\documentclass{PoS}

\pdfoutput=1 

\usepackage{journals}
\usepackage{url}
\usepackage[round]{natbib}
\shortcites{allam07,browne03,myers03,springel05}

\title{\href{http://www.astro.uni-bonn.de/~wucknitz/publications/pub.php?2007mru_poster}{Gravitational lens surveys with LOFAR}}

\ShortTitle{Gravitational lens surveys with LOFAR}

\author{\speaker{Olaf Wucknitz}\thanks{This work is supported by the European Community's Sixth Framework
Marie Curie Research Training Network Programme, Contract No.\
MRTN-CT-2004-505183 ``ANGLES'', and by the Emmy-Noether-Programme of the
`Deutsche Forschungsgemeinschaft', reference WU\,588/1-1.} \\
Argelander-Institut f\"ur Astronomie, Auf dem H\"ugel 71, 53121
Bonn, Germany \\
Joint Institute for VLBI in Europe, Oude Hoogeveensedijk 4, 7991 PD Dwingeloo,
The Netherlands \\
E-mail: \email{wucknitz@astro.uni-bonn.de}}

\author{M.~Garrett \\
ASTRON, Oude Hoogeveensedijk 4, 7991 PD Dwingeloo, The Netherlands}

\abstract{Deep surveys planned as a Key Science Project of LOFAR provide completely new 
opportunities for gravitational lens searches. For the first time do 
large-scale surveys reach the resolution required for a direct selection 
of lens candidates using morphological criteria.
We briefly describe the strategies that we will use to exploit this potential. 
The long baselines of an international \emph{E}-LOFAR are essential for this 
project.
}

\FullConference{From planets to dark energy: the modern radio universe\\
		 October 1-5 2007\\
		 University of Manchester, Manchester, UK}

\begin{document}

\section{Introduction}

Gravitational lensing is the best method to determine the mass distribution of
distant galaxies with high accuracy. With a large sample of lens systems at a
range of redshifts, we can study not only the structure
but also the evolution of galaxies.
We identify three main topics, in which lensing can help to resolve
controversial problems.
\begin{itemize}\parskip0pt\topsep0pt\partopsep0pt\parsep0pt\itemsep0pt
\item Most galaxies at moderate redshifts seem to have very close to
  isothermal ($\rho\propto r^{-2}$) mass distributions. Why is this the case,
  and does it also hold for higher redshifts?
\item How does the \emph{central} density profile of galaxies look like? Do
  they have cores or cusps, how do masses of central black holes evolve?
\item Are small sub-halos as abundant as predicted by the CDM structure
  formation scenario?
\end{itemize}

\begin{figure}[hb]
\hspace*{0.2\textwidth}%
\includegraphics[height=0.2\textwidth]{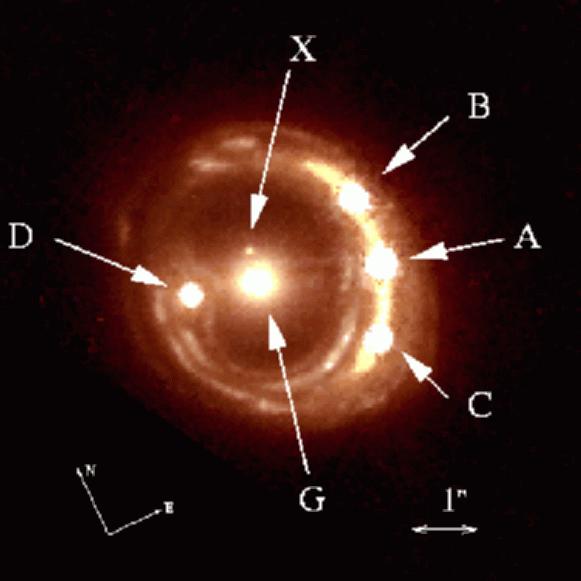}%
\hfill%
\includegraphics[height=0.2\textwidth]{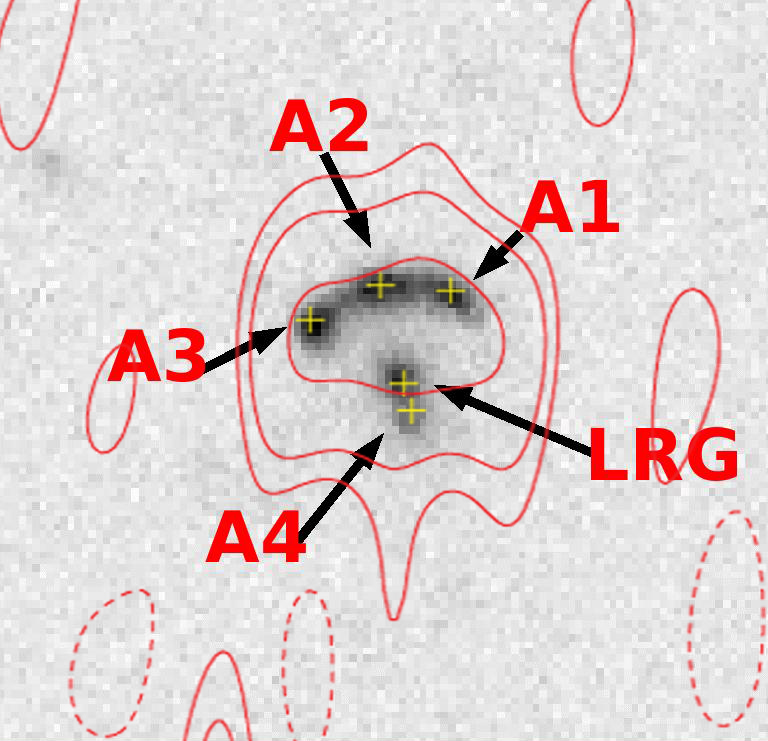}
\hspace*{0.2\textwidth}
\caption{Two examples from the bright end of typical lens systems expected to be
  found in LOFAR surveys.
  Left: RXS\,J1131--1231 \citep[HST image from][]{claeskens06}.
  Right: The `8 o'clock arc' \citep{allam07}. WSRT 5\,GHz contours superimposed
  upon SDSS image.
}
\label{fig:1131+8uur}
\end{figure}

We can help answering these questions by modelling the mass distributions to
fit for the structure of gravitationally lensed sources. In order to obtain
information about the complete mass distribution, lensed \emph{extended}
sources like the ones shown in Fig.~\ref{fig:1131+8uur} should be used and
modelled with LensClean \citep{lc1}.
There are several advantages in using radio observations.
Most important are the wide range
of resolutions that can be achieved with radio arrays and the absence of
effects like dust extinction and microlensing.
So far we know less than 50 radio lenses, whilst the number of optical lenses
has grown well above that and is increasing further.

\section{New lens surveys}

\subsection{Source-targeted search}

LOFAR not only probes a new frequency 
range but also new parameter space in terms of resolution and sensitivity of
wide-area radio surveys. With these data, it is for the first time possible
to identify lens systems directly from the source surveys using
morphological criteria.
Following up complete source samples as done in CLASS \citep{myers03,browne03}
would be prohibitive for the extreme number of sources expected in LOFAR
surveys. 

The expected lensing rates for this source population is 1:2000, leading to
potential numbers of 400\,000 and 15\,000 lenses in the LOFAR surveys at 120
and 200\,MHz, respectively.
Because of sensitivity and
resolution limitations, we will be able to identify only a small fraction of
these lenses. At 200\,MHz, \emph{900 lenses} are a more realistic
number, for which we probably have to follow-up around 10\,000 candidates with
the EVLA and/or \emph{e}-MERLIN. At
120\,MHz, longer baselines would be needed to detect more than just 
the systems with the largest image separations.

A good fraction of the LOFAR sources will be star-bursts galaxies, and the
majority will have extensions and structures exactly on the scale that is
needed for accuracte lens models. We can thus build the perfect sample for the
study of the structure and evolution of galaxies.

\subsection{Lens-targeted search}

An alternative strategy to find lens systems is to add the information from
optical galaxy catalogues listing potential lensing galaxies at moderate 
redshifts.
By cross-correlating these galaxy catalogues with the LOFAR source catalogues,
we can establish a large sample of lens systems with well-defined selection
criteria for the lens galaxies.
Combining the \emph{Luminous Red Galaxies} (LRG) from the SDSS with the
120\,MHz LOFAR survey, we expect to find about \emph{500 lenses} in 5000
candidates. With upcoming larger galaxy surveys, this number can be
significantly increased.

\section{Limitations}

\begin{figure}
\hspace*{0.02\textwidth}%
\rotatebox{90}{\qquad NL}~
\includegraphics[width=0.2\textwidth]{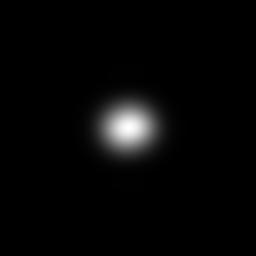}
\hfill%
\rotatebox{90}{\qquad NL + 5}~
\includegraphics[width=0.2\textwidth]{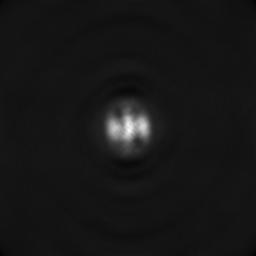}
\hfill%
\rotatebox{90}{\qquad NL + 16}~
\includegraphics[width=0.2\textwidth]{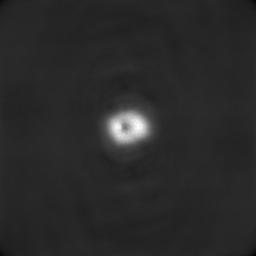}
\hspace*{0.02\textwidth}
\caption{Simulated Clean maps of a lensed star-burst galaxy with three
  different LOFAR arrays. The full international LOFAR is needed to properly
  resolve and map the lensed source.  }
\label{fig:maps}
\end{figure}

Even with additional stations in neighbouring countries, the resolution of
LOFAR will be sufficient to securely identify only lenses with relatively
large separations.
Fig.~\ref{fig:maps} shows simulations of a lensed star-burst galaxy observed
with three different arrays.
We see 
that a good number of international stations is not only required for
the resolution but also for good mapping properties.

The recently announced de-scope of LOFAR will modify the expectations for the
lens surveys. Very roughly, a drop in sensitivity by a factor of 2 will lead
to a reduction in the number of lenses by the same factor. Realistic
expectations will depend on the adapted survey strategy and the number of
international stations, which can partly compensate for the reduction of Dutch LOFAR.

\bibsep0.2ex
\bibliographystyle{aa}
\bibliography{poster}

\end{document}